# Least-action perihelion precession

Mikael Koskela[1] and Arto Annila[2,3,4,*]

[1]*Altrincham Grammar School for Boys, Bowdon, WA14 2RS, UK*
[2]*Department of Physics,* [3]*Institute of Biotechnology and* [4]*Department of Biosciences, FI-00014 University of Helsinki, Finland*

**Abstract**
The precession of Mercury's perihelion is reinspected by the principle of least action. The anomalous advancement of the apside line that is customarily accounted by the theory of general relativity, is ascribed to the gravitational effect due to the entire Universe. When the least action is written in the Sun's frame of reference, the residual rotation is seen to stem from inertia due to all bodies in the Universe. Since mass corresponds to a bound form of energy, gravity, as any other force, can be described as an energy density difference between a system of bodies and its surrounding energy densities that are dispersed throughout the Universe. According to the principle of least action the Universe is expanding by combustion of mass to radiation in the quest of equilibrating the bound forms of energy with "zero-density surroundings" in least time.

*Keywords: cosmological principle; energy density; energy dispersal; evolution; gravity; the principle of least action*

Theory of general relativity triumphed by providing a numerical value for the anomalous advancement of Mercury perihelion in an excellent agreement with observations (1,2). The residual rotation of the line of apsides was ascribed to the spacetime that curves about the Sun (3,4). However, for a layman the successful calculation, as such, does not disclose what exactly the curved spacetime means. Of course conceptual conundrums of general relativity do not only trouble the perihelion precession but involve other phenomena too, e.g., bending of light that is passing by a massive object (4,5) and ticking of a clock that is subject to acceleration (6) and increasing angular size of a galaxy that is ever further away from us (4,7). Yet the perihelion precession as a seemingly simple process prompts the layman to ask what will cause the planet to lengthen its path past the orbit's exact closure. On one hand when an expert answers by describing the stationary gravitational field of a star by the Schwarzschild metric (8), the layman's request to learn about the cause of the effect will not be satisfied. For him a causal connection would entail a flow time (9). On the other hand to look for some irreversible change does not parallel the thoughts of the expert – and the conversation relinquishes futile. But is the layman's call for comprehension by cause and effect all obsolete?

## Contributions to the least-action orbit

The Mercury's perihelion rotates 5600´´ per century in reference to Earth's equinox line (10). The rotation of our reference frame is 5025´´ and the gravitational tug of the other planets contributes 532´´ per century so that their sum 5557´´ falls short by 43´´ from the measured value (11). This anomalous part is according to the general relativity attributed to the spacetime that curves due to the presence of matter, here most notably due that of the Sun. Since the other planets also contribute, it seems logical for the layman to reason that the rest of the Universe must contribute to the precession too. Thus he regards the curved spacetime as an explanation that apparently sums up all these gravitational effects (Fig. 1). This line of reasoning is in a sense sound but it would be an unorthodox resolution since according to the general relativity, the curved spacetime, i.e., the metric gives rise to gravity (12). But is the layman's direct way of understanding all wrong?

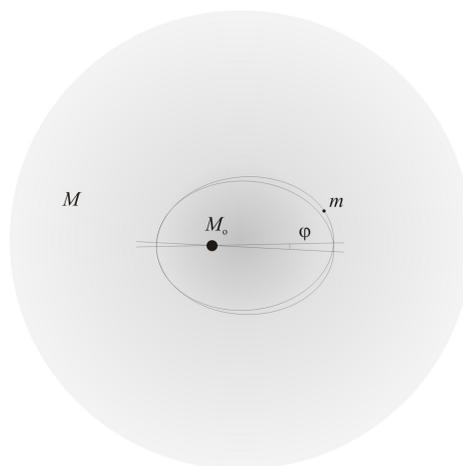

**Figure 1** Mercury ($m$) is on an elliptic orbit about the Sun ($M_o$). The line of apsides advances about the Sun at a gradual rate $d_t\varphi$. The gravitational tug of other planets contributes to the precession 532´´ per century whereas a residual of $\varphi = 43´´$ is described as a relativistic effect due to the curved spacetime. However, the layman reasons that, besides the Sun and its planets, the residual precession stems from the gravitational tug of all other bodies that total the mass $M$ of the Universe.

At first it may seem like a formidable task to account accurately for the total inertia due to some 100 billion stars in the Milky Way (13) as well as all those others in some 170 billion other galaxies that the Universe is estimated to house (14). Formally the inertia $I = \Sigma m_i r_i^2$ due to all masses $m_i$ at center-of-mass distances $r_i$

$$I = \sum_i m_i r_i^2 = \frac{\sum_{i<j} m_i m_j r_{ij}^2}{\sum_i m_i} = \frac{M_\circ m r^2}{M} + \cdots \quad (1)$$

can be rewritten in terms of distances $r_{ij}$ between any two bodies $m_i$ and $m_j$ in the Universe of total mass $M = \Sigma m_i$. In this way it becomes apparent that each and every pair of bodies is in relation to everything else in the Universe. The particular term involving the inertia $I = mr^2$ of the planet of mass $m$ orbiting at a distance $r$ from the Sun of mass $M_o$ is singled out from the huge series for further inspection.

The residual precession $\varphi$ can be identified by the principle of least action (15,16). The orbit is optimal when the kinetic energy $2K = I\omega^2$, given in terms of inertia $I$ and angular velocity $\omega$ is a constant integrand of the action $\int 2K dt$. In other words the action of the planet whose orbit advances a full precession period in $\tau = 1/\omega = 1/2\pi f = t/2\pi$ does not change

$$I\omega^2 = mr^2\omega^2 = (2\pi)^2 m \frac{v^2}{c^2} c^2 = \frac{2\pi^2 GM_\circ}{c^2 r} mc^2 \\ = (2\pi)^2 \frac{M_\circ}{r} \frac{R}{M} mc^2 = \varphi mc^2 \quad (2)$$

where the kinematic equation $r = \frac{1}{2}a_o t^2$ for the local acceleration $a_o = GM_o/r^2$ (17) is used to relate the planet's velocity $v = r/t$ to the mass $M_o$ via the constant of gravitation $G$. Likewise when $R = \frac{1}{2}aT^2$ for the universal acceleration $a = GM/R^2$ is used to relate the speed of light $c = R/T$ to the mass $M$ of the Universe that has expanded during time $T$ to the radius $R$, the result can also be stated so that the advancing arc $r\varphi$ of the perihelion's orbit about the Sun makes the fraction $M_o/M$ of the universal arc $R\Phi$. The obtained value $\varphi = 5.035 \cdot 10^{-7}$ rad, equivalent to 43.09˝ per century, is in agreement with observations $43.1 \pm 0.5˝$ (11). Numerical values of the Mercury's semi-major axis $r = 5.79 \cdot 10^{10}$ m and the Sun's mass $M_o = 1.989 \cdot 10^{30}$ kg (18) were used in Eq. 2. As usual, the semi-major axis $r$ of a circular or elliptical orbit relates via the Kepler's third law $t^2 = (2\pi)^2 r^3/GM_o$ to the orbital period $t$ of a small body of mass $m$ that revolves about the central body of mass $M_o$.

**The mass of the Universe**

The residual part of the perihelion precession in the universal surroundings (Eq. 2) can also be written as the energy ratio

$$\varphi = \frac{d_t L}{Q} = \frac{(2\pi)^2}{2} \frac{GmM_\circ}{r} \frac{1}{mc^2} = -\frac{(2\pi)^2}{2} \frac{U}{Q} \\ = (2\pi)^2 \frac{K}{Q} = (2\pi)^2 \frac{\frac{1}{2}mv^2}{mc^2} \quad (3)$$

of the Sun's gravitational potential $U = GmM_o/r$ and the planet's energy $Q = mc^2$ given in equivalents of radiation that the universal surroundings, i.e., the free space can absorb. Alternatively, when using the virial theorem $2K + U = 0$ of a stationary system, $\varphi$ can be expressed as the ratio of kinetic energy $K = \frac{1}{2}mv^2$ in the orbital motion and the dissipation $Q$. The energy ratio as a thermodynamic form is familiar to the layman since it relates the revolution rate of the planet to the energy in the surrounding in the same way as the maximum revs of an engine relate to temperature of the surroundings into which the engine exhausts its waste heat (19). Likewise a temperature gradient between a warm ocean and cold atmosphere powers a whirling hurricane (20).

All around hovering energy density (so called vacuum density) is characterized by the average temperature of the free space, at the moment $T$ = 2.725 K, as well as by permittivity and permeability that define the speed of light $c^2 = (\varepsilon_o \mu_o)^{-1}$ and the invariant impedance $Z^2 = \mu_o/\varepsilon_o$. The non-zero energy density manifests itself, e.g., in the Casimir effect (21) and Aharanov-Bohm experiment (22). Also the tiny but non-negligible universal acceleration $a = GM/R^2 = 1/\varepsilon_o\mu_o R = c^2/R = c/T$ reports from the energy density difference between energy that is still bound in the total mass $M$ of the Universe and "the zero-density surroundings". This energy gradient is diminishing at the rate $H = 1/t$. Today, after some $T$ = 13.7 billion years of diluting combustion, the radius $R(t)$ is huge $12.95 \cdot 10^{27}$ m. The rate $H = d_t a/a$ is changing as the scale factor (23) $a(t) = \sqrt{c^2 \varepsilon \mu}$ is stretching from the reference index of refraction $n = 1/a(T) = 1$ defined by the present-time permittivity $\varepsilon_o$ and permeability $\mu_o$. Ultimately, when combustion exhausts all repositories of bound energy, the expansion evens out the universal curvature to complete flatness.

The Universe's contribution to the perihelion precession is obtained by integrating the mass density $\rho$ from the present moment $t = 0$ back to $t = T$ that marks the nascent Universe. The density falls with time so that the more distant an object is, the less it will contribute to the Mercury's motion. Ultimately the nascent Universe that is receding from us at the speed of light will have no effect at all. At the largest scale $\rho$ is homogenous according to the cosmological principle. This can be understood as a consequence of the principle of least action. The least-time energy dispersal entails that diverse sources of radiation, i.e., galaxies containing gaseous clouds, planets, stars and black holes are dispersing from each other further and further apart so that the rate

$$d_t L = d_t(2Kt) = d_t(Mc^2 t) = d_t(Mr^2\omega) \\ = r^2 \omega d_t M + M d_t(r^2\omega) \quad (4)$$

is maximal, i.e., on the average the flows of energy distribute evenly (24). The non-conserved first term denotes by $d_t M$ the combustion of mass to radiation. It powers the expansion of the Universe. Thus, the mass of the Universe is not a constant. The conserved second term is familiar from the Kepler's second law where the specific relative angular momentum $r^2 d_t\theta$ is a constant of motion, e.g., so that the line from a planet to the Sun sweeps out equal areas during equal intervals of time of rotation $\omega = d_t\theta$. Since $d_t(r^2 d_t\theta)$ vanishes for a constant angular acceleration (25), the flows of energy will naturally select (26) the most effective means and mechanisms (27) in the quest of the maximal energy dispersal. Any accumulation of matter will even out via some mechanisms that transform mass to radiation. In other words the bound forms of energy are the loci of space. Also time is physical. It is a free form of energy that flows in transformations from a locus of space to another. At any moment of time and in any place of space the combustion $r^2\omega d_t M$ is maximal when on the average the mass density $\rho = 1/2\pi G t^2$ falls from any point inversely proportional to the square of the distance $r$. The least-time expansion ensures the uniform distribution of energy at the largest scale. The high precision of homogeneity manifests itself as the minute anisotropy in the cosmic background radiation (28). This is to say that each and every locus of space is to a high precision at a center of the Universe.

When $\rho$ is taken homogenous at all scales, the total mass of the expanding Universe is obtained by integration of the mass density from $r = 0$ to the radius $R = cT$

$$M = \int_0^R \rho 4\pi r^2 dr = \int_0^R \frac{1}{2\pi G t^2} 4\pi r^2 dr = \frac{2c^2 R}{G}. \quad (5)$$

Alternatively, one may use the observed perihelion precession of Mercury in Eq. 2 to compute the mass of the Universe $M = 3.488 \cdot 10^{53}$ kg.

The geometry of the expanding Universe can also be written so that $\rho$ relates to the area spanned by the angle of observation $\theta$ that contains the Universe at a past moment (Fig. 2). The integration is then over a spherical cap of height $h = r(1 - \cos(\theta/2)) = 2r\sin^2(\theta/4) = 2r/t^2$ up to the most distant past $T = R/c$

$$M = \int_0^R \frac{1}{2\pi G} 2\pi r h dr = \int_0^R \frac{1}{2\pi G} 4\pi r^2 \sin^2(\theta/4) dr \\ = \int_0^R \frac{4\pi r^2}{2\pi G t^2} dr = \int_0^T \frac{2c^3}{G} dt = \frac{2c^3 T}{G} \quad (6)$$

where $r\sin(\theta/4) = r'$ can be regarded as the radius of the Universe at a moment $t$ back in time. The opening angle $\theta$ is decreasing with $t$ and ultimately, when $t \to \infty$, $\theta \to 0$. For example, a distant object such as a galaxy, when viewed from earth, spans an arc $r\theta$ that is decreasing monotonically with $r = ct$. This is in contrast to Lambda cosmology where, objects with increasing redshift beyond about 1.5 would appear larger and larger (7,29). The light, that was emitted at a frequency $f_e$ to energy-dense surroundings at a moment $t_e$ back in time, distributes its energy on a longer and longer period, i.e., shifts red during its passage to the contemporary sparse-energy density. Thus, while the energy is conserved $2K = mv^2 = hf_o v^2/c^2 = hf_e$, the redshift $z = f_e/f_o - 1 = n^2 - 1$ reports via the decreasing index of refraction $n^2 = c^2/v^2 = \varepsilon\mu/\varepsilon_o\mu_o$ from the increasing radius of curvature due to the dilution of the energy density of free space. This understanding of the Doppler shift (30) is in agreement with Planck's law which says that on the average at any moment and at any place radiation and matter are in balance with each other.

The geometry of the expanding Universe can described so that the spherical cap spanned by the view angle has height $h = 2r/t^2 = 2c^2 r/r^2 = GM/r^2 = a$ which identifies to the acceleration which in turn relates to the radius of curvature $r = \frac{1}{2}at^2$ at time $t$. In other words, when the Universe was young, $r$ was smaller and accordingly $a$ was stronger than it is today. The irreversible combustion of matter to photons $d_t M/M = d_t r/r = H = 1/t$, as given by Eq. 6, generates the universal arrow of time (9). When the bound forms of energy, i.e., matter breaks down, photons as free forms of energy will escape forever. The force of expansion is by today $t$ = 13.7 billion years very weak since the mass density $\rho = 1/2\pi G t^2$ is very low 12.78·10$^{-27}$ kg/m$^3$. This calculated value though is higher than the estimate 9.9·10$^{-27}$ kg/m$^3$ obtained from the Wilkinson Microwave Anisotropy Probe (WMAP) measurements (31). However the WMAP data has been interpreted on the basis of Friedmann-Lemaître-Robertson-Walker (FLRW) metric. Therefore the critical density $\rho_c = 3H^2/8\pi G$ of the Friedman equation is lower by the geometrical factor $3/4$ than provided by the formula $\rho = H^2/2\pi G$ (32,33,34).

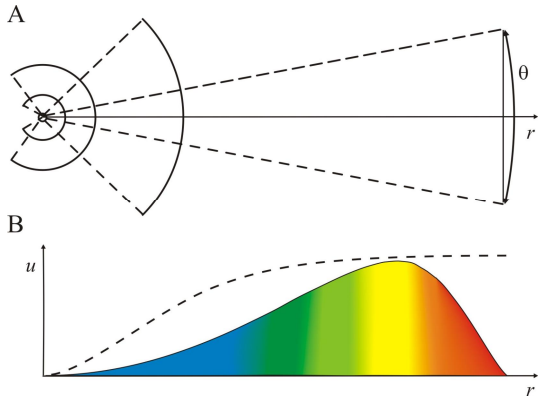

**Figure 2** (A) Geometry of the expanding Universe is such that the area spanned by the angle of observation $\theta$ when viewed from any locus of space contains the Universe at the moment $t$ back in time at the radius $r = ct$. The present moment $t$ = 0 and $r$ = 0 corresponds to $\theta = 2\pi$ whereas the nascent Universe at $T$ = 13.7 billion years back of radius $R = cT$ corresponds to a tiny angle $\theta$ = 2.3·10$^{-18}$ rad. Today the same spot is seen in every direction. Nothing can be observed from this ultra deep field spot that is receding from us at the speed of light. Since the huge Universe is to a good approximation flat, a small arc $r\theta$ can be approximated by a chord of Euclidean geometry according to the Pythagorean theorem $r(1 - v^2/c^2)^{1/2}$. (B) On the average the energy density $u(r)$ (solid line) accumulates (dashed line) with increasing $r$ about any locus in the Universe so the few brightest (blue) bodies are nearby whereas numerous faint (red) bodies are further away. Ultimately the spectral density vanishes altogether at $R$. The distribution displays the balance between spectral energy density and energy density of mass according to the Planck's law.

**Gravitational force as an energy density difference**
Although the above elementary calculation reveals that the anomalous part of the perihelion precession stems via inertia from the total mass of the Universe, the layman's quest to understand the phenomenon via cause and effect remains to be clarified. In other words gravity as any other force must be described so that it exerts its effect via some flow of energy. In particular the force carrier of causality remains to be identified.

It is well-known that for a pair of opposite charges $q^+$ and $q^-$ the resulting electromagnetic field $-\nabla\Phi$ vanishes on the dipole axis. Although there is no electromagnetic field on the dipole axis the potential energy density $\Phi \propto 1/r$ does exist. Thus there is energy density about a net neutral body that comprises of numerous dipoles in random orientations. Since wavelengths of standing waves are limited by the distance between boundaries, not all those standing modes of energy densities that propagate in the universal surroundings (35) will fit between any two bodies. Therefore the energy density between the two bodies is different from that in the surroundings. The gravitational force is this energy density difference. It is an attractive force when the energy density with the system of bodies is higher than that of its surroundings (Fig. 3). When the distance between the two bodies decreases, energy is expelled to the surroundings. However the energy density between the two bodies will increase because the accommodated wavelengths will decrease. In other words increasingly higher frequencies are only allowed when bodies move closer and closer to each other. Consequently the potential energy density gradient $-\nabla U \propto 1/r^2$ relative to the surroundings, i.e., the force will increase quadratic with decreasing distance $r$.

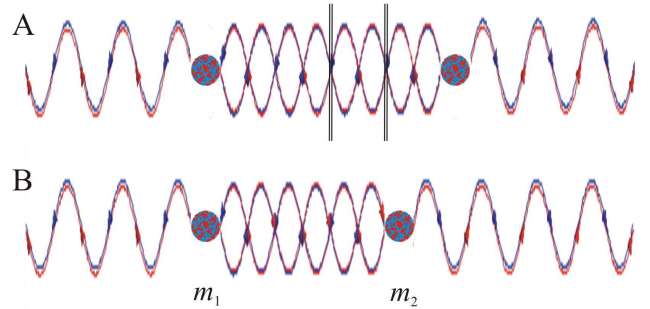

**Figure 3**. (A) Gravitational force arises from the energy density difference between the two net neutral bodies and their surroundings. The energy density about a body is exemplified by a propagating wave that stems from the net neutral aggregates of pairs of opposite charges (red and blue) that accumulate the mass $m_1$ and $m_2$. For clarity the decay of the energy density $\propto 1/r$ is omitted. (B) Energy-sparse surroundings cause attraction by accepting energy from the bound modes that pair the two bodies along gravitational field lines. The force is consumed in changes of state when energy flows from the bound system to its surroundings in quanta that are carried by the doubly paired density modules of photons with opposite phases. These are marked as ||∞|| and are known as gravitons.

The fundamental modulus of a standing wave of energy density, i.e., a gravitational wave is a pair of bosons that have opposite phases since they have been generated by pairs of opposite charges. The standing wave can be expressed as a sum of oppositely propagating waves subject to boundary conditions (36). This balance between radiation and matter is also familiar from the derivation of Planck's law where a set of discrete states is approximated by a continuum to obtain a continuous distribution of energy densities (25). Since the modulus of density wave between net neutral bodies is without a longitudinal vector character, a traceless tensor (cvf. electromagnetic wave) denotes transverse modes of the force carrier, known as the graviton.

When gravity is understood as the force due to energy density differences, then the residual in the perihelion precession is understood to reflect ultimately the energy density difference between the Universe and its "zero-density surroundings". The force is decreasing irreversibly by flows of energy from high to low densities that bring about the increasing universal radius of curvature. This universal force, often described as the curvature of spacetime, is the cause of the perihelion precession anomaly as requested by the layman.

**Discussion**

The layman's calculation of the residual in the Mercury perihelion precession by the principle of least action applies likewise for other planets in the solar system (11), the asteroid Icarus (37) and binary pulsar systems (38). When the obtained functional form $\varphi = 2\pi^2 GM_o/c^2r$ is compared with the approximate relativistic formula $\varphi \approx 6\pi GM_o/c^2r$, the two can be seen to differ from each other as much as 6 differs from $2\pi \approx 6.283$. The corresponding difference to the full relativistic calculation is negligible. Moreover, the layman calculation gives a meaning to the relativistic calculation by relating the Schwarzschild radius $r_s$ of the Sun, about 3 km, with the radius $R$ of the Universe. In other words, it is not the mass $M_o$ of an object that defines the radius of a horizon $r_h = GM_o/2c^2 = RM_o/M$ but it is the ratio of the local energy density to the surrounding energy density contained in the mass $M$ of the Universe. Consequently, when the Universe was young and dense, $r_h$ for a given mass $M_o$ was smaller than in today's sparse surroundings. More importantly calculations based on the principle of least action, unlike those based merely on energy, are not troubled by a singularity $r \to 0$ because the absolutely least action, the quantum of action $h$ is finite (39). This is to say that the symmetry group U(1) that characterizes electromagnetic radiation cannot be broken down.

Furthermore, the account of the evolving Universe as a natural process that breaks irreversibly from one stationary state of symmetry (40) to another of different symmetry is revealing. Since the Poincaré group is the full symmetry group of any relativistic field theory (41), also the general relativity cannot but describe a stationary state (42) and thus it fails to account accurately for the series of state changes that the expanding Universe is undergoing.

The layman's holistic view of gravity parallels Mach's thinking of inertia that nothing can be done without affecting everything else (43,44,45). Gravity as a force is the manifestation of energy density differences that couple everything to everything else via flows of energy. The space is connected, i.e., affine via these flows of energy and finite but without boundary. Since there are numerous pathways for the force carriers to propagate and to diminish the energy density differences among all bound forms of energy, the Universe is expanding uniformly in the least time.

A view of sky picks a connected 3-manifold (e.g., a spherical cap) that is homeomorphic to the 3-sphere at a moment back in time (Eqs. 5 and 6). It is natural to perceive the space as three-dimensional because a stationary state's least-action path is on a plane spanned by two orthogonal eigenvectors and defined by the normal of the plane. Time is the fourth dimension that associates with a change of state where a confined circulation transforms to another by excising quanta from the bound state to the surroundings or including free quanta from the surroundings to the system. Evolution of space, as it was emphasized above, is not a continuous process but it proceeds in steps of quantized actions from one state of symmetry to another. Thus, a loop of space cannot be continuously tightened to a point as was conjectured by Poincaré (46) but ultimately the spontaneous symmetry breaking processes attains the absolutely least closed circulation. When it breaks open, the resulting open action belongs to the absolutely lowest group of symmetry U(1) which is the symmetry group of electromagnetic radiation (47,39).

The holistic portrayal of the perihelion precession by the principle of least action is self-consistent. Yet it may appear ancient and perhaps to some even a fortuitous coincidence when giving the numerical value in agreement with observations. However for the layman the explanation is not an *ad hoc* resolution since it is based on the same natural principle that has been used to clarify diverse phenomena extending from quarks to galaxies and from biology to economics (48,49,50,51,52,53,54,55).

**Acknowledgements**

We thank Jani Anttila and Aki Kallonen for insightful comments.

**References**

1. Einstein, A. 1916 The foundation of the general theory of relativity. *Annalen der Physik* **49**, 769–822.
2. Chow, T. L. 2008 *Gravity, black holes, and the very early universe: an introduction to general relativity and cosmology*. New York, NY: Springer.
3. Berry, M. 1974 *Principles of cosmology and gravitation*. Cambridge, UK: Cambridge University Press.
4. Narlikar, J. V. 1993 *Introduction to cosmology*. Cambridge, UK: Cambridge University Press.
5. Will, C. M. 2006 The confrontation between general relativity and experiment. *Living Rev. Relativity* **9**, 3. http://www.livingreviews.org/lrr-2006-3.
6. Resnick, R. 1968 *Introduction to special relativity*. New York, NY: John Wiley & Sons.
7. Raine, D. J. & Thomas, E. G. 2001 *An introduction to the science of cosmology*. Bristol, UK: Institute of Physics.
8. Weinberg, S. 1972 *Gravitation and cosmology: Principles and applications of the general theory of relativity*. New York, NY: John Wiley & Sons.
9. Tuisku, P., Pernu, T. K. & Annila, A. 2009 In the light of time. *Proc. R. Soc. A.* **465**, 1173–1198.
10. Brown, K. 1999 *Reflections on relativity*. Mathpages.com
11. Clemence, G. M. 1947 The relativity effect in planetary motions. *Rev. Mod. Phys.* **19**, 361–364.
12. Wheeler, J. A. 1990 *A journey into gravity and spacetime*. Scientific American Library, San Francisco: W. H. Freeman.
13. http://www.esa.int/esaSC/SEM75BS1VED_index_0.html.
14. Gott, III, J. R., et al. 2005 A map of the universe. *Astrophys. J.* **624**, 463–484.
15. De Maupertuis, P. L. M. 1746 Les Loix du Mouvement et du Repos Déduites d'un Principe Metaphysique. *Mém. Ac. Berlin* 267–294.
16. Kaila, V. R. I. & Annila, A. 2008 Natural selection for least action. *Proc. R. Soc. A.* **464**, 3055–3070.
17. Breithaupt, J. 2008 *Physics A*. Cheltenham, UK: Nelson Thornes.
18. The JPL HORIZONS on-line solar system data http://ssd.jpl.nasa.gov/?horizons.
19. Kittel, C. 1969 *Thermal physics*. New York, NY: W. H. Freeman Company.
20. Emanuel, K. A. 1994 *Atmospheric convection*. New York, NY: Oxford University Press.
21. Casimir, H. B. G. & Polder, D. 1948 The influence of retardation on the London-van der Waals forces. *Phys. Rev.* **73**, 360–372.
22. Aharonov, Y. & Bohm, D. 1959 Significance of electromagnetic potentials in quantum theory. *Phys. Rev.* **115**, 485–491.
23. Weinberg, S. 1972 *Gravitation and cosmology, principles and applications of the general theory of relativity*. New York, NY: Wiley.
24. Annila, A. 2009 Space, time and machines. (arXiv:0910.2629).
25. Alonso, M. & Finn, E. J. 1983 *Fundamental university physics*. Reading, MA: Addison-Wesley.
26. Darwin, C. 1859 *On the origin of species*. London, UK: John Murray.
27. Sharma, V. & Annila, A. 2007 Natural process – Natural selection. *Biophys. Chem.* **127**, 123–128.
28. Wright, E. L. 2004 Theoretical Overview of Cosmic Microwave Background Anisotropy, in *Measuring and modeling the Universe*. Ed. Freedman, W. L. Carnegie Observatories Astrophysics Series. Cambridge, MA: Cambridge University Press.
29. Mattig, W. 1958 Über den Zusammenhang zwischen Rotverschiebung und scheinbarer Helligkeit. *Astron. Nach.* **284**, 109–111.
30. Suntola, T. 2009 *The Dynamic Universe, Toward a unified picture of physical reality*. Physics Foundations Society.
31. Bennett, C. L. et al., 2003 First-year Wilkinson Microwave Anisotropy Probe (WMAP) Observations: Preliminary maps and basic results. *Astrophys. J. Suppl. Series* **148**, 1–27.
32. Lemaître, G. 1927 Un univers homogène de masse constante et de rayon croissant rendant compte de la vitesse radiale des nébuleuses extra-galactiques. *Annales de la Société Scientifique de Bruxelles* **47**, 49–56.
33. Hubble, E. 1929 A relation between distance and radial velocity among extra-galactic nebulae. *Proc. Natl. Acad. Sci. USA* **15**, 168–173.
34. Unsöld, A. & Baschek, B. 2002 *The new cosmos, an introduction to astronomy and astrophysics*. New York, NY: Springer-Verlag.
35. Casimir, H. B. G. & Polder, D. 1948 The influence of retardation on the London-van der Waals forces. *Phys. Rev.* **73**, 360–372.
36. Crawford, F. S. 1968 *Waves* in *Berkeley physics course*; 3 New York, NY: McGraw-Hill.
37. Shapiro, I. I., Ash, M. E., Smith, W. B. 1968 Icarus: Further confirmation of the relativistic perihelion precession. *Phys. Rev. Lett.* **20**, 1517–1518.

38. Kramer, M. et al. 2006 Tests of general relativity from timing the double pulsar. *Science* **314**, 97–102.
39. Annila A. All in action (arXiv:1005.3854)
40. Noether, E. 1918 Invariante Variationprobleme. *Nach. v.d. Ges. d. Wiss zu Goettingen, Mathphys. Klasse* 235–257; English translation Tavel, M. A. 1971 Invariant variation problem. *Transp. Theory Stat. Phys.* **1**, 183–207.
41. Weinberg, S. 1995 *The Quantum theory of fields*. Cambridge, UK: Cambridge University press.
42. Birkhoff, G. D. 1924 *Relativity and Modern Physics*. Cambridge, MA: Harvard University Press..
43. Bondi, H. & Samuel, J. 1996 The Lense–Thirring effect and Mach's principle. (arXiv:9607009).
44. Von Bayer, H. C. 2001 *The Fermi solution: Essays on science*. Mineola, NY: Dover Publications.
45. Einstein A. Letter to Ernst Mach, Zürich, 25 June 1923, in Misner, C., Thorne, K. S., Wheeler, J. A. 1973. *Gravitation*. San Francisco, CA: W. H. Freeman.
46. http://www.claymath.org/millennium/Poincare_Conjecture/
47. Griffiths, D. 1999 *Introduction to electrodynamics*. Englewood Cliffs, NJ: Prentice Hall.
48. Feynman, R. P. 1948 Space-time approach to non-relativistic quantum mechanics. *Rev. Mod. Phys.* **20**, 367–387.
49. Bak, P. 1996 *How nature works: The science of self-organized criticality*. New York, NY: Copernicus.
50. Lineweaver, C. H. & Egan, C. A. 2008 Life, Gravity and the Second Law of Thermodynamics. *Physics of Life Reviews* **5**, 225–242.
51. Georgiev, G. & Georgiev, I. 2002 The least action and the metric of an organized system. *Open Systems and Information Dynamics* **9**, 371–380.
52. Annila, A. & Salthe, S. N. 2010 Physical foundations of evolutionary theory. *J. Non-equil. Thermodyn.* **35**.
53. Annila, A. & Salthe, S. 2009 Economies evolve by energy dispersal. *Entropy* **11**, 606–633.
54. Salthe, S. N. 1985 *Evolving hierarchical systems: Their structure and representation*. New York, NY: Columbia University Press.
55. Beeson, D. 1992 *Maupertuis: an intellectual biography*. Oxford, UK: Voltaire Foundation.